\documentclass[letterpaper,preprintnumbers,english,floats,floatfix,amssymb,prd,superscriptaddress,twocolumn,nofootinbib]{revtex4}
\usepackage{amssymb,amsmath}
\usepackage{epsfig,hyperref}
\usepackage[svgnames]{xcolor}
\usepackage{pgf,tikz}
\usepackage{epstopdf}
\usepackage[british]{babel}
\usepackage{enumerate}
\usepackage{enumitem}
\usepackage{makecell}
\usepackage{booktabs}
\usepackage{graphicx}
\usepackage{nicefrac}
\usepackage{units}
\usepackage{mathrsfs}
\usepackage{placeins} %

\makeatletter
\DeclareRobustCommand*{\bfseries}{%
	\not@math@alphabet\bfseries\mathbf
	\fontseries\bfdefault\selectfont
	\boldmath
}
\makeatother

\def\be{\begin{equation}}
\def\ee{\end{equation}}
\def\beq{\begin{eqnarray}}
\def\eeq{\end{eqnarray}}

\makeatletter

\newcommand{\arXiv}[2][]{\href{http://arxiv.org/abs/#2}{\texttt{arXiv:#2\@ifempty{#1}{}{ [#1]}}}}
\makeatother

\begin{document}
\title{Critical collapse of a massive scalar field in semi-classical loop quantum gravity}
	
\author{Li-Jie Xin}
\affiliation{School of Physics and Optoelectronics, South China University of Technology, Guangzhou 510641, China}
	
	
\author{Xiangdong Zhang\textsuperscript{1}}
\email{ scxdzhang@scut.edu.cn}
	
\date{\today}
	
\begin{abstract}
We investigate critical phenomena during the gravitational collapse of a massive scalar field under two distinct semi-classical loop quantum gravity (LQG) approaches within spherical symmetry. Numerical simulations reveal that the massive scalar field in both semi-classical frameworks exhibits two distinct types of critical behavior, consistent with the classical scenario. When the scalar field's mass parameter is small, type II critical phenomena emerge, with the resulting echoing periods and critical exponents precisely matching those obtained in general relativity. In contrast, a large mass parameter triggers type I critical phenomena, where the resulting black holes possess a finite minimum mass. These findings suggest that semi-classical corrections from LQG have a negligible impact on the dynamics of critical collapse.

\end{abstract}
\maketitle

\section{Introduction\label{sec:introduction}}
Gravitational collapse occurs when a massive celestial body collapses inward due to its internal pressure being unable to resist its own gravity. Critical gravitational collapse examines the dynamics near the threshold between black hole formation and dispersal to infinity. Choptuik pioneered research into this critical phenomenon~\cite{Choptuik:1992jv}. In his study of the spherical gravitational collapse of a massless scalar field, he identified a critical threshold for black hole formation, denoted as 
$p^{*}$, within a one-parameter family of initial data. When the initial parameter exceeds this critical threshold ($p > p^{*}$), the scalar field collapses to form a black hole. Conversely, when the initial parameter is below this critical threshold ($p < p^{*}$), the scalar field escapes to infinity, leaving behind a flat spacetime. Near the critical threshold ($p \sim p^{*}$), critical phenomena emerge. Choptuik found that critical gravitational collapse exhibits three key features: universality, self-similarity, and power law scaling. Universality implies that, near the critical threshold, all initial configurations remain close to the critical solution for a finite duration. Self-similarity indicates that the evolution of spacetime geometry and matter fields lacks characteristic scales and displays an exact, periodic scaling behavior, characterized by an echoing period $\Delta \approx 3.4$ that is independent of the initial data. Power law scaling states that, in the super-critical regime, the mass of the black hole follows the power law relation $M_{BH} \propto |p-p*|^{\gamma}$, where the critical exponent $\gamma \approx 0.37$ is universal and independent of the initial data.According to this scaling, black holes have no minimum mass; by fine-tuning the initial parameter, one can produce black holes with arbitrarily small mass. Subsequently, Hod and Piran~\cite{Hod:1996az} discovered that the power law scaling of black holes also exhibits a fine structure. This fine structure manifests as a periodic function of the critical parameter ($p-p^{*}$), and can be expressed as $\ln{M_{BH}} \approx \gamma\ln{|p-p*|}+c_{f}+\Psi\left[\ln{|p-p*|}\right]$, where $\Psi\left[\ln{|p-p*|}\right]$ denotes a small periodic correction term.

In Ref.~\cite{Choptuik:1996yg}, Choptuik  \emph{et al.} performed numerical simulations of the gravitational collapse of spherically symmetric $SU(2)$ Yang-Mills fields and identified two distinct critical behaviors at the black hole formation threshold. The first type of critical behavior is analogous to the previously studied critical solution for a massless scalar field: it exhibits discrete self-similarity and permits the formation of black holes with arbitrarily small masses, corresponding to a type II critical phenomenon. In contrast, the second type lacks the power law scaling of black hole masses; instead, the resulting black holes possess a finite lower bound in mass, characteristic of a type I critical phenomenon.

In Ref.~\cite{Brady:1997fj}, Brady \emph{et al.} investigated the critical gravitational collapse of a massive real scalar field, uncovering two distinct types of critical behavior. They found that type I critical collapse occurs when the characteristic scale of the scalar field exceeds $1/m$, whereas type II critical collapse emerges when the characteristic scale is smaller than $1/m$, where $m$ denotes the mass of the scalar field. The critical gravitational collapse of a massive scalar field can be regarded as a simplified analog of the $SU(2)$ Yang-Mills case. Ref.~\cite{Jimenez-Vazquez:2022fix} further explored the critical gravitational collapse of a massive complex scalar field, identifying critical exponents consistent with Choptuik's findings and two distinct types of critical phenomena. Subsequent studies have extended this analysis to diverse matter fields~\cite{Hirschmann:1994du,Brady:1994aq,Frolov:1999fv,Singh:1994tb,Koike:1999eg,Neilsen:1998qc,Carr:1999xs,Gundlach:1999cw,Gundlach:1996je,Choptuik:1996yg,Choptuik:1999gh,Bizon:2001ju}, various modified gravity theories~\cite{Eardley:1995ns,Husa:2000kr, Ventrella:2003fu,Golod:2012yt}, and axisymmetric configurations~\cite{Abrahams:1993wa,Ledvinka:2021rve, Baumgarte:2023tdh}. Comprehensive reviews on the critical gravitational collapse can be found in Refs.~\cite{Gundlach:1999cu,Gundlach:2002sx}.

Inspired by these developments, a key question arises: how do quantum gravity effects influence the collapsing process of scalar fields? Among quantum gravity approaches, loop quantum gravity (LQG) stands out due to its background independence and non-perturbative nature \cite{Ashtekar:2004eh, Rovelli 2004, Thiemann, Han:2005km,Zhang:2023yps}. Benítez \emph{et al.} investigated the critical gravitational collapse of a spherically symmetric, minimally coupled massless scalar field within semi-classical LQG~\cite{Benitez:2020szx}. Their results revealed that the system retains the exact scale invariance of its classical counterpart, yielding a critical exponent identical to Choptuik's result. In a follow-up study~\cite{Benitez:2021zjs}, they demonstrated that this semi-classical system also preserves the \textquoteleft wiggles\textquoteright{} in the power law scaling of black hole masses, akin to the fine structure observed in classical general relativity.  Gambini \emph{et al.} proposed a covariant polymerization procedure for a spherically symmetric massless scalar field coupled to gravity~\cite{Gambini:2021uzf}, obtaining results consistent with classical general relativity.

Considering these advancements, this work aimed to investigate the gravitational collapse of a massive real scalar field within two semi-classical LQG frameworks~\cite{Benitez:2020szx,Gambini:2021uzf}. Specifically, we examine whether the introduction of a scalar field potential triggers critical behavior that deviates from the established classical system..

The paper is organized as follows. In Sec.~\ref{sec:methodology}, we detail the methodology for deriving the dynamical equations under two distinct polymerization procedures for scalar variables. Secs.~\ref{sec:case1} and \ref{sec:case2} are devoted to discussing the results of critical gravitational collapse obtained within each procedure. Finally, a summary and discussion of our key findings are presented in Sec.~\ref{sec:summary}. Throughout this paper, geometric units are adopted such that $G=c=1$.

\section{Methodology\label{sec:methodology}}
\subsection{Polymerization procedure I}
Initially, following the approach in Ref~\cite{Benitez:2020szx}, we construct the semi-classical equations of motion by polymerizing the scalar field according to the prescription $\phi \to\frac{\sin\left(k\varphi\right)}{k}$. The classical gravitational variables are represented by the spherical remnants of the triads in the radial and transverse directions, denoted as $E^{x}$ and $E^{\varphi}_{1}$ along with their canonically conjugate momenta, $K_{x}$ and $K_{\varphi}$. The spacetime metric can then be expressed as:
\be
ds^{2} = -N_{1}^{2}\left(t,r\right) dt^{2}+\Lambda^{2}\left(t,r\right) dr^{2}+R^{2}d\Omega^{2}, \label{line_element1}
\ee
where the metric components are related to the LQG triads via $\Lambda = E^{\varphi}_{1}/r$ and $R^{2} = |E^{x}|$. The extrinsic curvature components are given by $K_{xx} = -\text{sign}(E^{x})(E^{\varphi}_{1})^{2}K_{x}/ \sqrt{|E^{x}|}$ and $K_{\theta\theta}=-\sqrt{|E^{x}|}K_{\varphi}/(2\beta)$, where $\beta$ denotes the Immirzi parameter. We adopt the gauge $E^{x} = r^{2}$, which effectively eliminates the $K_{x}$ component by solving the diffeomorphism constraint. Furthermore, the polar slicing condition in these variables corresponds to setting $K_{\varphi} = 0$.

The classical Einstein equations and the equation of motion for the massive scalar field $\phi$ are given by:
\be
\frac{N_{1}'}{N_{1}} - \frac{(E^{\varphi}_{1})'}{E^{\varphi}_{1}} + \frac{2}{r} - \frac{(E^{\varphi}_{1})^{2}}{r^{3}} + 8\pi\frac{(E^{\varphi}_{1})^{2}}{r}V\left(\phi\right) = 0 ,
\label{class_alpha}
\ee
\be
\frac{(E^{\varphi}_{1})'}{E^{\varphi}_{1}} - \frac{3}{2r}+\frac{(E^{\varphi}_{1})^{2}}{2 r^{3}} - 2\pi r(\Pi^{2}+\Phi^{2}) - 4\pi\frac{(E^{\varphi}_{1})^{2}}{r}V\left(\phi\right)=0,
\label{class_Ephi}
\ee
\be
\dot{\Phi} = \left(\frac{N_{1} r}{E^{\varphi}_{1}}\Pi\right)' ,
\label{class_Phi}
\ee
\be
\begin{split}
&\dot{\Pi} = \frac{rN_{1}}{E^{\varphi}_{1}}\Phi'  - rN_{1} E^{\varphi}_{1} \frac{dV\left(\phi\right)}{d\phi} \\
&+\left(\frac{rN_{1}'}{E^{\varphi}_{1}} +\frac{3N_{1}}{E^{\varphi}_{1}} -\frac{rN_{1}\left(E^{\varphi}_{1}\right)'}{\left(E^{\varphi}_{1}\right)^{2}}\right)\Phi,
\end{split}
\label{class_Pi}
\ee
where $\Phi =\phi'$, $\Pi = \frac{E^{\varphi}_{1}}{N_{1} r}\dot{\phi}$, and the potential term is $V\left(\phi\right)=\mu^{2}\phi^{2}/2$. Here, primes and overdots denote derivatives with respect to $r$ and $t$, respectively.

In the context of the polymerized scalar field, the transition to the semi-classical equations is achieved through the replacements $\phi \to \frac{\sin\left(k\varphi\right)}{k}$ and $P_{\phi} \to P_{\varphi}$, where $k$ is the polymerization parameter. The resulting modified equations of motion are: 
\be
\frac{N_{1}'}{N_{1}} - \frac{(E^{\varphi}_{1})'}{E^{\varphi}_{1}} + \frac{2}{r} - \frac{(E^{\varphi}_{1})^{2}}{r^{3}} + 4\pi\mu^{2}\frac{\left(E^{\varphi}_{1}\right)^{2}\sin^{2}\left(k\varphi\right)}{r k^{2}} = 0,
\label{case1_alpha}
\ee
\begin{equation}
\begin{split}
    \frac{(E^{\varphi}_{1})'}{E^{\varphi}_{1}} - \frac{3}{2r}+&\frac{(E^{\varphi}_{1})^{2}}{2 r^{3}} - 2\pi r\left(\frac{\left(P_{\varphi}\right)^{2}}{r^{4}}+\left(\varphi'\right)^{2}\cos^{2}\left(k\varphi\right)\right) \\
&- 2\pi\mu^{2}\frac{(E^{\varphi}_{1})^{2}\sin^{2}\left(k\varphi\right)}{r k^{2}} =0,
\end{split}
\label{case1_Ephi}
\end{equation}
\be
\dot{\varphi} = \frac{N_{1}}{E^{\varphi}_{1}r}P_{\varphi},
\label{case1_psi}
\ee
\be
\begin{split}
    \dot{P_{\varphi}} = &\frac{r^{2}}{E^{\varphi}_{1}}\left[\left(\frac{3N_{1} E^{\varphi}_{1}-rN_{1}\left(E^{\varphi}_{1}\right)'+N_{1}'E^{\varphi}_{1}r}{E^{\varphi}_{1}}\right)\varphi'\cos^{2}\left(k\varphi\right)\right.  \\
    &\left.+rN_{1}\varphi''\cos^{2}\left(k\varphi\right)-rN_{1} k \left(\varphi'\right)^{2}\cos\left(k\varphi\right)\sin\left(k\varphi\right) \right]  \\
    &-r\mu^{2}N_{1} E^{\varphi}_{1} \frac{\sin\left(k\varphi\right)}{k}.
\end{split}
\label{case1_Pi}
\ee

The classical equations of motion are recovered in the limit $k \to 0$. To ensure regularity at the origin, we impose the boundary conditions $N_{1}(t,0) = 1$ and $\varphi'(t,0) =0$. The initial profile of the scalar field is given by: 
\be
\varphi \left(0,r\right) = p \exp\left[\frac{\left(r-r_{0}\right)^{2}}{\sigma^{2}}\right],
\label{initial_psi}
\ee

\be
P_{\varphi}\left(0,r\right) = 0.
\label{initial_Pi}
\ee
where $p$, $r_{0}$, and $\sigma$ represent the amplitude, the center position, and the width of the initial scalar field, respectively.

The numerical algorithm is implemented as follows: initially, the profiles for the scalar field $\varphi$ and its conjugate momentum $P_{\varphi}$ are specified. The metric functions $N_{1}$ and $E^{\varphi}_{1}$ are then determined by integrating the constraint equations (\ref{case1_alpha}) and (\ref{case1_Ephi}), using a fourth-order Runge-Kutta method. Subsequently, the scalar field variables $\varphi$ and $P_{\varphi}$ are advanced to the next time step by integrating the evolution equations (\ref{case1_psi}) and (\ref{case1_Pi}) via the fourth-order Runge-Kutta scheme. The metric field data at this new time step are determined by integrating Equations (\ref{case1_alpha}) and (\ref{case1_Ephi}) again. This process is repeated iteratively to obtain the metric and scalar field data throughout the entire spacetime.

To monitor black hole formation, we utilize the Misner-Sharp mass, which represents the total energy enclosed within a sphere of radius $r$,
\be
\begin{split}
m&\equiv\frac{r}{2}(1-g^{\mu\nu}r_{,\mu}r_{,\nu})\\
&=\frac{r}{2}\left(1-\frac{r^{2}}{\left(E^{\varphi}_{1}\right)^{2}}\right).\\
\end{split}
\label{MS}
\ee
The formation of a black hole is numerically identified when the ratio $2m/r$ exceeds the threshold of $0.9$, at which point the evolution is terminated.

\subsection{Polymerization procedure II}
This subsection follows the polymerization procedure for the scalar field as presented in Ref.~\cite{Gambini:2021uzf}. The total Hamiltonian is given by:
\begin{equation}
\begin{split}
    H_{T} = &\int dx \bigg\{ N^{x} \left( (E^{x})' K_{x} - E^{\varphi} (K_{\varphi})' - 8\pi P_{\phi} \phi' \right) \\ 
    &+ N \left[ -\frac{E^{\varphi}}{2\sqrt{|E^{x}|}} - 2\sqrt{|E^{x}|} K_{\varphi} K_{x} - \frac{K^{2}_{\varphi} E^{\varphi}}{2\sqrt{|E^{x}|}} \right. \\ 
    &+ \frac{((E^{x})')^{2}}{8\sqrt{|E^{x}|} E^{\varphi}} - \frac{\sqrt{|E^{x}|} (E^{x})' (E^{\varphi})'}{2(E^{\varphi})^{2}} + \frac{\sqrt{|E^{x}|} (E^{x})''}{2E^{\varphi}} \\
    &+ \frac{2\pi P^{2}_{\phi}}{\sqrt{|E^{x}|} E^{\varphi}} + \frac{2\pi (|E^{x}|)^{3/2} (\phi')^{2}}{E^{\varphi}} \\
    &\left. + 4\pi E^{\varphi} \sqrt{|E^{x}|} V(\phi) \right] \bigg\}. 
\end{split}
\label{hamiltonian}
\end{equation}

The potential $V\left(\phi\right)=\mu^{2}\phi^{2}/2$ represents the mass term of the scalar field. By redefining the shift vector and lapse function as: 
\be
\overline{N^{x}} = N^{x} + \frac{2NK_{\varphi}\sqrt{|E^{x}|}}{\left(E^{x}\right)'},
\ee
\be
\overline{N} = \frac{NE^{\varphi}}{\left(E^{x}\right)'}.
\ee
The total Hamiltonian can be rewritten as
\be
H_{T} = \int dx \left(\overline{N^{x}}D_{x}+\overline{N}H\right),
\ee
where $D_{x}$ is the diffeomorphism constraint:
\be
D_{x} = \left(E^{x}\right)'K_{x} - E^{\varphi}\left(K_{\varphi}\right)'-8\pi P_{\phi}\phi',
\ee
and $H$ is the redefined Hamiltonian constraint,
\be
\begin{split}
    H = &\left[\sqrt{|E^{x}|}\left(\frac{\left(\left(E^{x}\right)'\right)^{2}}{4\left(E^{\varphi}\right)^{2}}-1-K_{\varphi}^{2}\right)\right]'- \frac{2K_{\varphi}\sqrt{|E^{x}|}\phi'P_{\phi}}{E^{\varphi}} \\
    &+\frac{2\pi\left(E^{x}\right)'P_{\phi}^{2}}{\sqrt{|E^{x}|}\left(E^{\varphi}\right)^{2}}+\frac{2\pi\left(|E^{x}|\right)^{3/2}\left(E^{x}\right)'\left(\phi'\right)^{2}}{\left(E^{\varphi}\right)^{2}} \\
    &+2\pi \sqrt{|E^{x}|}\left(E^{x}\right)'\mu^{2}\phi^{2}).
\end{split}
\ee
We apply the following canonical transformations to the scalar field $\phi$, the curvature $K_{\varphi}$, and their canonical momenta $P_{\phi}$ and $E^{\varphi}$,
\be
\phi \to \frac{\sin\left(k\varphi\right)}{k},    P_{\phi} \to \frac{P_{\varphi}}{\cos\left(k\varphi\right)},
\ee
\be
K_{\varphi} \to \frac{\sin\left(\rho K_{\varphi}\right)}{\rho},    E^{\varphi} \to \frac{E^{\varphi}}{\cos\left(\rho K_{\varphi}\right)},
\ee
where $k$ and $\rho$ are the polymerization parameters. Substituting these transformations, we obtain:
\begin{equation}
    \begin{split}
       & H =  \frac{\sqrt{E^{x}}\left(E^{x}\right)'\left(E^{x}\right)''\cos^{2}\left(\rho K_{\varphi}\right)}{2\left(E^{\varphi}\right)^{2}} + \frac{\left(\left(E^{x}\right)'\right)^{3}\cos^{2}\left(\rho K_{\varphi}\right)}{8\left(E^{\varphi}\right)^{2}\sqrt{E^{x}}} \\
        & -\frac{\sqrt{E^{x}}\left(\left(E^{x}\right)'\right)^{2}\left(E^{\varphi}\right)'\cos^{2}\left(\rho K_{\varphi}\right)}{2\left(E^{\varphi}\right)^{3}} \\
        &- \frac{\sqrt{E^{x}}\left(\left(E^{x}\right)'\right)^{2}\rho\left(K_{\varphi}\right)'\sin\left(\rho K_{\varphi}\right)\cos\left(\rho K_{\varphi}\right)}{2\left(E^{\varphi}\right)^{2}}\\
        &-\frac{\left(E^{x}\right)'}{2\sqrt{E^{x}}}  \left(1+\frac{\sin^{2}\left(\rho K_{\varphi}\right)}{\rho^{2}}\right) - \frac{2\sqrt{E^{x}}\sin\left(\rho K_{\varphi}\right)\cos\left(\rho K_{\varphi}\right)\left(K_{\varphi}\right)'}{\rho} \\
        &- \frac{2\sin\left(\rho K_{\varphi}\right)\cos\left(\rho K_{\varphi}\right)\sqrt{E^{x}}P_{\varphi}\varphi'}{\rho E^{\varphi}} + \frac{2\pi\left(E^{x}\right)'\cos^{2}\left(\rho K_{\varphi}\right)P_{\varphi}^{2}}{\sqrt{E^{x}}\left(E^{\varphi}\right)^{2}\cos^{2}\left(k\varphi\right)} \\
        &+\frac{2\pi\sqrt{E^{x}}E^{x}\left(E^{x}\right)'\left(\varphi'\right)^{2}\cos^{2}\left(k\varphi\right)\cos^{2}\left(\rho K_{\varphi}\right)}{\left(E^{\varphi}\right)^{2}} \\
        & +\frac{2 \pi \sqrt{E^{x}}\left(E^{x}\right)'\mu^{2}\sin^{2}\left(k\varphi\right)}{k^{2}}.
    \end{split}
    \label{ham2}
\end{equation}
Under the gauge choice $E^{x}=x^{2}$, $K_{\varphi}=0$, the semi-classical equations of motion are derived as:
\be
\frac{N'}{N}-\frac{\left(E^{\varphi}\right)'}{E^{\varphi}}+\frac{2}{x}-\frac{\left(E^{\varphi}\right)^{2}}{x^{3}}+\frac{4\pi\mu^{2} \left(E^{\varphi}\right)^{2}\sin^{2}\left(k\varphi\right)}{xk^{2}}=0,
\label{shift2}
\ee
\be
\begin{split}
    &\frac{\left(E^{\varphi}\right)'}{E^{\varphi}}-\frac{3}{2x}+\frac{\left(E^{\varphi}\right)^{2}}{2x^{3}}-\frac{2\pi\mu^{2}\left(E^{\varphi}\right)^{2}\sin^{2}{\left(k\varphi\right)}}{xk^{2}} \\
&-2\pi x \left(\frac{\left(P_{\varphi}\right)^{2}}{x^{4}\cos^{2}\left(k\varphi\right)}+\left(\varphi'\right)^{2}\cos^{2}\left(k\varphi\right)\right)=0,
\end{split}
\label{lapse2}
\ee
\be
\dot{\varphi}=\frac{4\pi N P_{\varphi}}{E^{\varphi}x\cos^{2}\left(k\varphi\right)},
\label{psi2}
\ee

\begin{equation}
\begin{split}
    \dot{P_{\varphi}} = &-\frac{4\pi N k P_{\varphi}^{2}\sin(k\varphi)}{E^{\varphi}x\cos^{3}(k\varphi)} + \frac{4\pi x^{2}}{E^{\varphi}} \bigg[ xN\varphi''\cos^{2}(k\varphi) \\
    &+\left(\frac{3NE^{\varphi}-xN(E^{\varphi})'+N'xE^{\varphi}}{E^{\varphi}}\right)\varphi'\cos^{2}(k\varphi) \\
    &-xNk(\varphi')^{2}\cos(k\varphi)\sin(k\varphi) \bigg] \\
    &- \frac{4\pi \mu^{2}x N E^{\varphi} \cos(k\varphi)\sin(k\varphi)}{k}.
\end{split}
\label{pi2}
\end{equation}

\begin{figure*}[t!]
	\centering
	\begin{tabular}{cc}
		\includegraphics[width=0.85\textwidth]{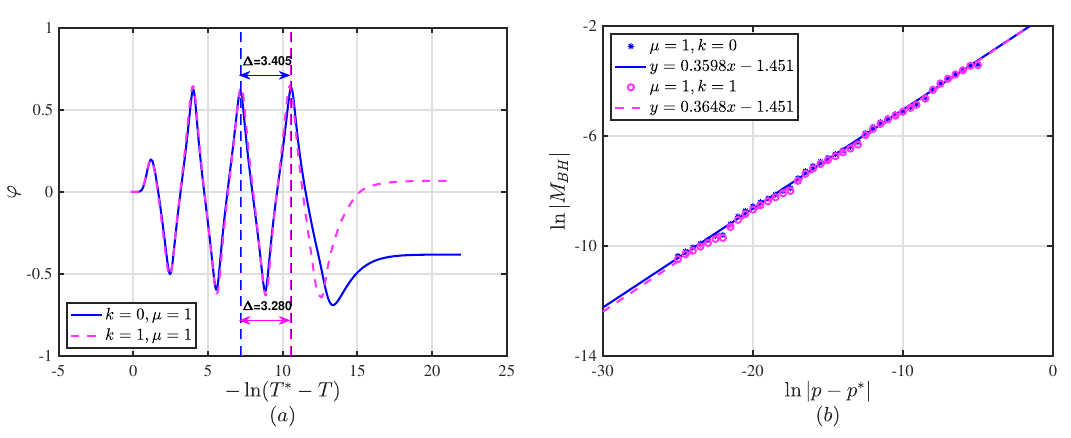}
	\end{tabular}
	\caption{Critical behavior of a massive scalar field with potential $V\left(\phi\right) = \mu^{2}\phi^{2}/2$ for $\mu=1.0$. (a) Evolution of the scalar field $\varphi$ at the origin $(r = 0)$ as a function of $-\ln{(T^{*}-T})$, where $T^{*}$ denotes the time of naked singularity formation.  (b) Power-law scaling of the black hole mass in the supercritical regime. The initial profile parameters in Equation~(\ref{initial_psi}) are $r_{0}=1.0$, and $\sigma=0.2$. The blue solid line and the magenta dashed line correspond to the cases $k=0$ and $k=1$, respectively.}
	\label{fig:prlslop}
\end{figure*}

The classical equations of motion are recovered in the limit $k \to 0$. To ensure regularity at the origin ($x=0$), we impose the boundary conditions $N(t,0)=1$ and $\varphi'(t,0)=0$. The initial conditions for the scalar field are specified as 
\be
\varphi(0,x) = p \exp\left[\frac{\left(x-x_{0}\right)^{2}}{\sigma^{2}}\right],
\label{initial_psi2}
\ee

\be
P_{\varphi}(0,x)  = 0.
\label{initial_Pi2}
\ee

\begin{table}[htb]
	\centering
	\begin{tabular}{c|c|c}    
		\hline
		\rule{0pt}{1.5em}Initial condition  & Echoing period($\Delta$) &  Critical exponent($\gamma$) \\
		\hline
		\rule{0pt}{1.5em}	$k=0,\mu=1.0$ &$3.405$   &$0.3598$  \\
		\hline
		\rule{0pt}{1.5em}	$k=1,\mu=1.0$  &$3.280$   &$0.3648$  \\
		\hline
		\rule{0pt}{1.5em}	$k=0,\mu=2.0$   &$3.367$   &$0.3616 $  \\
		\hline
        \rule{0pt}{1.5em}	$k=1,\mu=2.0$   &$3.442$   &$0.3616 $  \\
        \hline
        \rule{0pt}{1.5em}	$k=0,\mu=3.0$   &$3.528$   &$0.3417 $  \\
        \hline
        \rule{0pt}{1.5em}	$k=1,\mu=3.0$   &$3.329$   &$0.3581 $  \\
        \hline
        
	\end{tabular}
\caption{The echoing period $\Delta$ (second column) and critical exponent $\gamma$ (third column) for the collapse of massive scalar fields with various mass parameters are presented. The initial condition in Equation~(\ref{initial_psi}) is set to $r_{0}=1.0$ and $\sigma = 0.2$. (Both the echoing period and critical exponent are presented with four significant figures.)}
	\label{tab:case1}
\end{table}
The numerical strategy employed in this subsection is identical to that used in the previous subsection. Specifically, we use the fourth-order Runge-Kutta method to integrate the metric functions and evolve the equation of the scalar field. The Misner-Sharp energy in this subsection is also described by Eq. (\ref{MS}). Furthermore, the criterion for black hole formation remains the same as in the preceding subsection.

\begin{figure}[t!]
	\centering
	\begin{tabular}{cc}
		\includegraphics[width=0.45\textwidth]{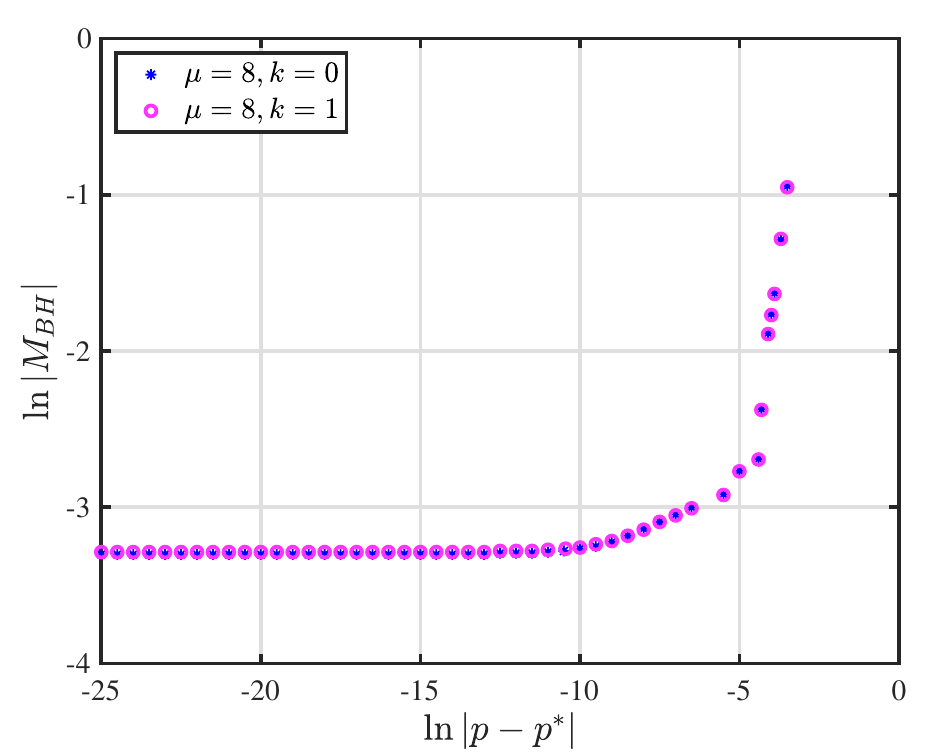}
	\end{tabular}
	\caption{Relationship between the black hole mass and the initial amplitude in the supercritical regime for a scalar field with potential $ V \left(\phi \right) = \mu^{2}\phi^{2}/2 $ and mass parameter $\mu = 8.0$. The blue stars and magenta circles correspond to the cases $k=0$ and $k=1$, respectively.}
	\label{fig:prltype}
\end{figure}

\section{Result I: polymerization procedure I\label{sec:case1}}

We first perform numerical simulations of the semi-classical, spherically symmetric massive scalar field described by Equations (\ref{case1_alpha})-(\ref{case1_Pi}) under the polymerization procedure I. To accurately extract the echoing period, a mesh refinement algorithm is implemented~\cite{Zhang:2016kzg, Hu:2023qcq}. By fixing the initial parameters at $r_{0}=1.0$ and $\sigma = 0.2$, we vary the amplitude of the initial scalar field $p$ to locate the critical threshold $p^{*}$. Our numerical results confirm the existence of critical behavior: if the amplitude is below the threshold $p^{*}$, the field eventually disperses with no black hole formation; conversely, if it exceeds the threshold $p^{*}$, gravitational collapse inevitably leads to the formation of a black hole.

\begin{figure*}[t!]
	\centering
	\begin{tabular}{cc}
		\includegraphics[width=0.85\textwidth]{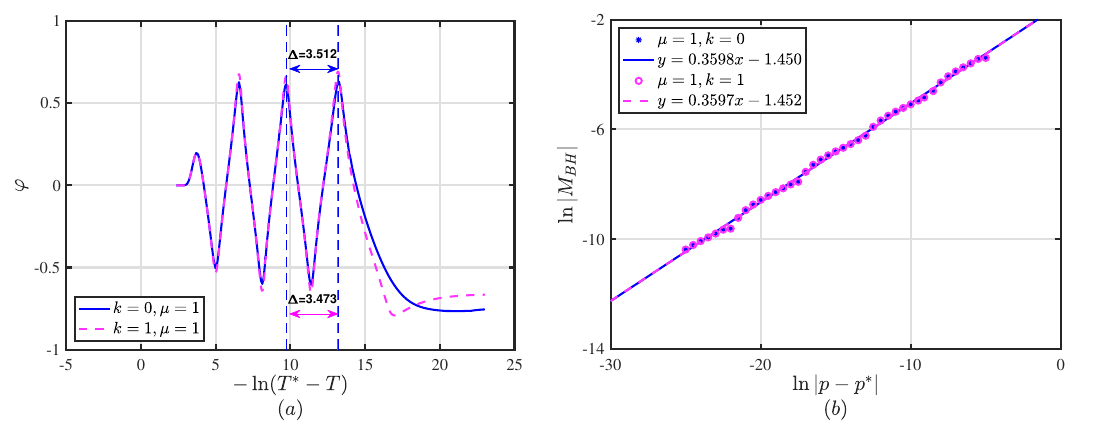}
	\end{tabular}
	\caption{Critical behavior of a massive scalar field with potential $V\left(\phi\right) = \mu^{2}\phi^{2}/2$ for $\mu=1.0$. (a) Evolution of the scalar field $\varphi$ at the origin $(x = 0)$ as a function of $-\ln{(T^{*}-T)}$, where $T^{*}$ denotes the time of naked singularity formation.  (b) Power-law scaling of the black hole mass in the supercritical regime. The initial profile parameters in Equation~(\ref{initial_psi2}) are $x_{0}=1.0$, and $\sigma=0.2$. The blue solid line and the magenta dashed line correspond to the cases $k=0$ and $k=1$, respectively.}
	\label{fig:unvslop}
\end{figure*}

\begin{figure}[t!]
	\centering
	\begin{tabular}{cc}
		\includegraphics[width=0.45\textwidth]{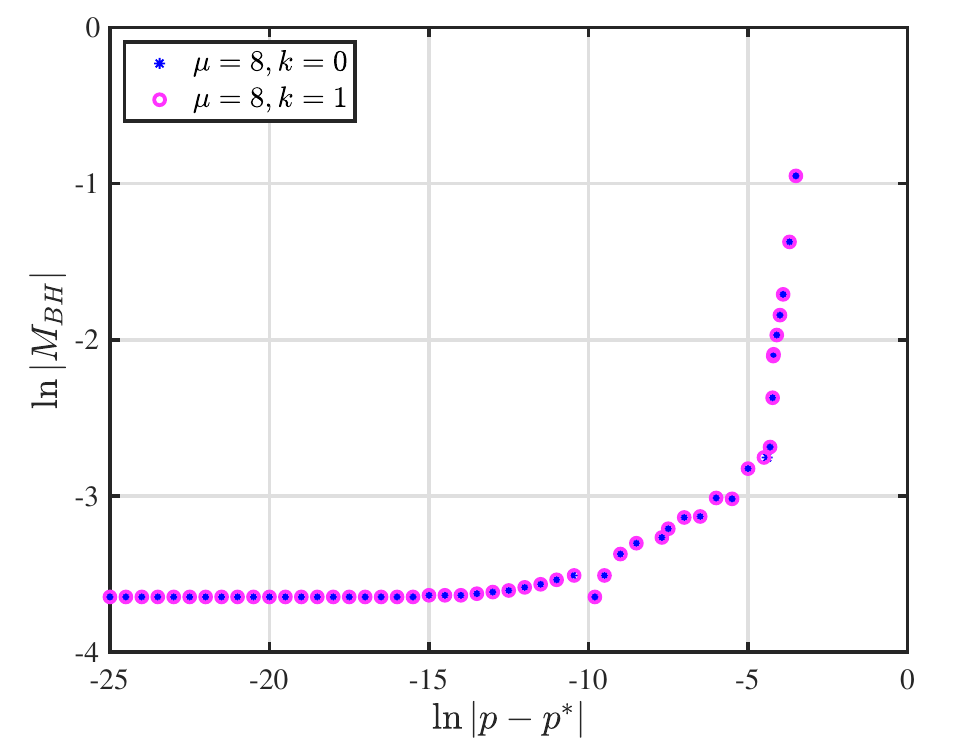}
	\end{tabular}
	\caption{Relationship between the black hole mass and the initial amplitude in the supercritical regime for a scalar field with potential $ V \left(\phi \right) = \mu^{2}\phi^{2}/2 $ and mass parameter $\mu = 8.0$. The blue stars and magenta circles correspond to the cases $k=0$ and $k=1$, respectively.}
	\label{fig:unvtype}
\end{figure}

Figure~\ref{fig:prlslop} (a) illustrates the evolution of the scalar field at the center as a function of $-\ln{(T^{*}-T)}$, where $T^{*}$ denotes the time of naked singularity formation. For a massive scalar field with $\mu=1.0$, the echoing period is found to be approximately $3.4$, a value that remains robust regardless of the inclusion of semi-classical corrections. This result is in excellent agreement with Choptuik’s results~\cite{Choptuik:1992jv}. Figure~\ref{fig:prlslop} (b) demonstrates the power law scaling of the black hole mass for various amplitudes $p$ in the supercritical regime. The results indicate that the critical exponent remains approximately $0.37$, with or without semi-classical corrections, further consistent with Choptuik’s results.

We further investigate the critical gravitational collapse of massive scalar fields with mass parameters $\mu=2.0$ and $\mu=3.0$. The results reveal that the echoing period and the critical exponent consistently remain near $3.4$ and $0.37$, respectively. The corresponding numerical data are summarized in Table~\ref{tab:case1}.

It is well established that the nature of the collapse depends on the product of the mass parameter and the characteristic scale of the scalar field: a product less than unity leads to type II critical behavior, while a product greater than unity leads to type I critical behavior~\cite{Brady:1997fj}. To examine the impact of  semi-classical corrections on type I critical behavior, we simulate a scalar field with $\mu=8.0$ while maintaining the same initial profile. As shown in Figure~\ref{fig:prltype}, in the supercritical regime, the black hole mass exhibits a finite lower bound regardless of the presence of semi-classical corrections.

\section{Result II: polymerization procedure II\label{sec:case2}}

We perform numerical simulations of the semi-classical, spherically symmetric massive scalar field described by Equations (\ref{shift2})–(\ref{pi2}) under the polymerization procedure II. To precisely determine the echoing period, we employ a mesh refinement algorithm as in the previous subsection. We fix the initial position and width of the scalar field at $x_{0}=1.0$ and $\sigma=0.2$, respectively, and vary the amplitude $p$ to locate the critical threshold $p^{*}$. For a mass parameter $\mu=1.0$, the numerical results indicate the existence of a critical threshold for the initial amplitude: when the amplitude is below this value, no black hole forms, whereas when it exceeds this value, gravitational collapse leads to the formation of a black hole.

\begin{table}[htb]
	\centering
	\begin{tabular}{c|c|c}    
		\hline
		\rule{0pt}{1.5em}Initial condition  & Echoing period($\Delta$) &  Critical exponent($\gamma$) \\
		\hline
		\rule{0pt}{1.5em}	$k=0,\mu=1.0$ &$3.512$   &$0.3598$  \\
		\hline
		\rule{0pt}{1.5em}	$k=1,\mu=1.0$  &$3.437$   &$0.3597$  \\
		\hline
		\rule{0pt}{1.5em}	$k=0,\mu=2.0$   &$3.440$   &$0.3638 $  \\
		\hline
        \rule{0pt}{1.5em}	$k=1,\mu=2.0$   &$3.334$   &$0.3604 $  \\
        \hline
        \rule{0pt}{1.5em}	$k=0,\mu=3.0$   &$3.232$   &$0.3412 $  \\
        \hline
        \rule{0pt}{1.5em}	$k=1,\mu=3.0$   &$3.220$   &$0.3438 $  \\
        \hline
        
	\end{tabular}
\caption{The echoing period $\Delta$ (second column) and critical exponent $\gamma$ (third column) for the collapse of massive scalar fields with various mass parameters are presented. The initial condition in Equation~(\ref{initial_psi2}) is set to $x_{0}=1.0$ and $\sigma = 0.2$. (Both the echoing period and critical exponent are presented with four significant figures.)}
	\label{tab:case2}
\end{table}

Figure~\ref{fig:unvslop} (a) shows the evolution of the scalar field at the center with respect to $-\ln{(T^{*}-T)}$, where $T^{*}$ denotes the time of naked singularity formation. This figure reveals that the echoing period for $\mu=1.0$ is approximately $3.4$ and remains unaffected by semi-classical corrections. Figure~\ref{fig:unvslop} (b) presents the variation of black hole mass relative to the initial amplitude in the supercritical regime, from which a critical exponent of approximately $0.37$ is derived, consistent with Choptuik’s results. Subsequently, we investigate the critical gravitational collapse of massive scalar fields with mass parameters of $\mu=2.0$ and $\mu=3.0$, obtaining the same echoing periods and critical exponents as those found in the $\mu=1.0$ case; specific numerical values are listed in Table~\ref{tab:case2}.

To examine the impact of semi-classical corrections on type I critical behavior in a more covariant LQG framework, we fix the initial position and width of the scalar field at $x_{0}=1.0$ and $\sigma=0.2$, respectively, and perform numerical simulations for a mass parameter  of $\mu=8.0$. In the supercritical regime, the relationship between the black hole mass and the initial scalar field amplitude is shown in Figure~\ref{fig:unvtype}. The results indicate that the semi-classical corrections do not affect type I critical behavior, and the formed black hole mass possesses a finite lower bound.

\section{Summary\label{sec:summary}}
We have investigated the critical collapse of a massive scalar field within two distinct semi-classical, spherically symmetric Loop Quantum Gravity (LQG) frameworks and reached two primary conclusions.

In the first semi-classical scheme, where modifications are confined to the scalar field sector, we observe prominent type II critical phenomena in the regime of a small mass term. The resulting echoing periods and mass scaling laws exhibit high consistency with the predictions of general relativity. Notably, no mass gap is present, indicating the absence of a minimum threshold for black hole formation. Conversely, in the regime of a large mass term, the collapse exhibits characteristic type I critical behavior, mirroring the results of general relativity. Here, the power-law scaling is replaced by a finite mass gap, and the resulting black hole possesses a non-zero minimum mass. Our results indicate that the polymerization parameter $k$ exerts no observable influence on these critical phenomena.

Subsequently, we examine a more covariant semi-classical scheme. Consistent with the previous framework, our numerical simulations reveal two distinct types of critical collapse behavior and confirm that the polymerization parameter $k$ does not affect the critical phenomena.

\section*{Acknowledgments}\small
This work is supported by National Natural Science Foundation of China (NSFC) with Grants No.12275087.

\FloatBarrier	

\end{document}